\documentclass[10pt,twocolumn,letterpaper]{article}

\usepackage[pagenumbers]{cvpr} 

\usepackage{graphicx}
\usepackage{amsmath}
\usepackage{xcolor}
\usepackage{colortbl}
\usepackage{amssymb}
\usepackage{booktabs}

\usepackage[pagebackref,breaklinks,colorlinks]{hyperref}

\usepackage[capitalize]{cleveref}
\crefname{section}{Sec.}{Secs.}
\Crefname{section}{Section}{Sections}
\Crefname{table}{Table}{Tables}
\crefname{table}{Tab.}{Tabs.}

\usepackage{bbding}
\usepackage{xcolor}
\definecolor{green}{RGB}{0,255,0}
\definecolor{red}{RGB}{255,0,0}
\usepackage{multirow}
\usepackage{stfloats}
\usepackage{bm}
\def\cvprPaperID{8383} 
\def\confName{CVPR}
\def\confYear{2023}

\begin{document}

\title{JoyTTS: LLM-based Spoken Chatbot With Voice Cloning}

\author{
    Fangru Zhou,
    Jun Zhao,
    Guoxin Wang\\
    JD Health International Inc. \\
}
\maketitle

\begin{abstract}
  JoyTTS is an end-to-end spoken chatbot that combines large language models (LLM) with text-to-speech (TTS) technology, featuring voice cloning capabilities. This project is built upon the open-source MiniCPM-o and CosyVoice2 models and trained on 2000 hours of conversational data. We have also provided the complete training code to facilitate further development and optimization by the community. On the testing machine seed-tts-zh, it achieves a SS (speaker similarity) score of 0.73 and a WER (Word Error Rate) of 5.09. The code and models, along with training and inference scripts, are available at \url{https://github.com/jdh-algo/JoyTTS.git}. 
\end{abstract}

\section{Methods}
    \subsection{Overview}
    
        Several end-to-end chatbot models exist, such as Qwen2.5-Omni\cite{Qwen2.5-Omni} and Lama-Omni2\cite{LLaMA-Omni2}. While these models provide robust conversational capabilities, they do not support voice cloning and lack open-source training code.
        
        MiniCPM-o\cite{yao2024minicpm} is an open source large language model designed for efficient text generation and understanding. Initially developed as an end-to-end chatbot, MiniCPM-o included a GPT-Sovits-based TTS module. However, the voice cloning capabilities of this initial TTS model were limited, which prompted further enhancements.

        To improve the voice cloning performance, the TTS component of MiniCPM-o was replaced with CosyVoice2\cite{CosyVoice2}. CosyVoice2 offers advanced speech synthesis capabilities and significantly enhances the naturalness and accuracy of voice cloning in JoyTTS.

    \subsection{JoyTTS model}
        
        \begin{figure*}[!htb]
        \centering
        \includegraphics[width=0.8\linewidth]{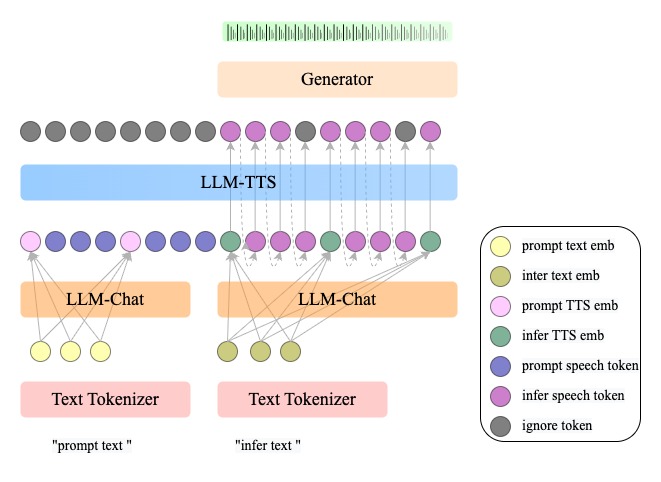}
        \caption{The framework of our LLM-based Spoken Chatbot model JoyTTS with voice cloning.}
        \label{fig:1}
        \end{figure*}

        As illustrated in Figure \ref{fig:1}, we introduce the JoyTTS model, which combines large language models (LLM) with text-to-speech (TTS) technology, featuring voice cloning capabilities.JoyTTS consists of four primary modules:

        Tokenizer Module: This module converts both text and audio into tokens, which are then fed into the LLM-Chat module.

        LLM-Chat Module: We employ the MiniCPM-o's LLM model, specifically the Qwen-7B model, as the LLM-Chat module. This module processes the input tokens and produces text outputs along with hidden layer features. These features are crucial for providing the LLM-TTS module with a deeper understanding of the contextual semantics of the text.

        LLM-TTS Module: Using the outputs from the LLM-Chat module, including the hidden layer features, the LLM-TTS module generates speech tokens. The inclusion of hidden layer features helps the LLM-TTS module better grasp the context and semantics of the conversation.

        Generator Module: This module takes the speech tokens produced by the LLM-TTS module and generates the final output, which includes the Mel spectrogram and audio.

\begin{figure*}[!htb]
        \centering
        \includegraphics[width=0.5\linewidth]{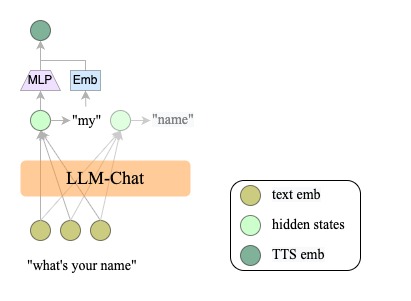}
        \caption{Diagram of TTS Embedding Generation.}
        \label{fig:2}
        \end{figure*}

         As illustrated in Figure \ref{fig:2}, transform the prompt text embeddings through LLM-Chat to produce hidden states. Then hidden states ( $h_{i}$) are mapped from 3584 dimensions to 768 dimensions through MLP layer and embedding layer. These mapped features are then combined with text embeddings ( $Emb(y_{i})$ ) to form the TTS embed, which is input into the LLM-TTS module.
        \begin{equation}
        \begin{aligned}
        TTS_{embed} = Emb(y_{i})+MLP(h_{i})
        \end{aligned}
        \end{equation}
        where $y_{i}$ is the text token corresponding to $h_{i}$ output by the LLM-Chat module.
                
        During inference, both the prompt text and prompt wave are pre-processed by the LLM-Chat module to obtain TTS embed, which is used as prior knowledge input to the LLM-TTS module, facilitating effective voice cloning.The input text of the LLM-Chat module is "Please repeat the following text. $text_{prompt}$"
       
\section{Data Construction}
    Our data contains 400K multi-turn text dialog samples,approximately 2k hours in total. These samples are derived from two open-source datasets: RedGPT\cite{redgpt} and GeneratedChat0.4M\cite{GeneratedChat} . To convert text dialogues into audio, we employed CosyVoice2. To enhance the model's voice cloning capabilities, we randomly selected text-audio pairs from the WenetSpeech4TTS dataset during audio conversion, using these as prompt text and prompt wav.

    To enhance the training dataset, we employed two data augmentation techniques. First, we split the dialogue texts into shorter segments of varying lengths, which helps the model learn to handle diverse conversational structures and adapt to different text lengths. Second, we introduced special punctuation marks into the text. This addition aims to improve the model's ability to interpret and generate nuanced speech patterns, reflecting more natural and expressive dialogue. These augmentations collectively contribute to a more robust and versatile model.
    
\section{Training}
     
    The training process for our model is structured into two distinct stages, each designed to progressively build and refine the capabilities of the system.

    In the first stage, we focus on the independent training of the LLM-Chat and LLM-TTS modules. The LLM-Chat module is prioritized to be fully trained initially to ensure that its hidden states and text labels are precisely aligned. This alignment is crucial as it forms the foundation for accurate text representation and understanding, which are essential for generating coherent and contextually relevant dialogue. Concurrently, the LLM-TTS module is trained to convert text into speech, focusing on achieving high-quality audio output that is both natural and expressive. By training these modules separately, we allow each to specialize and optimize its respective functions without interference, setting a solid groundwork for integration.
    
    In the second stage, we integrate the LLM-Chat and LLM-TTS modules for joint training. This phase involves optimizing the combined system using a unified loss function, which balances the objectives of both modules. The joint training allows for the fine-tuning of interactions between text generation and speech synthesis, ensuring that the dialogue produced is not only contextually appropriate but also delivered in a natural and engaging manner. By optimizing the combined loss, we enhance the model's overall performance, achieving seamless integration between text understanding and voice generation capabilities.

    Throughout both stages, careful attention is paid to maintaining the quality of voice cloning, ensuring that the model can replicate diverse vocal characteristics with high fidelity. This structured approach to training not only improves the individual components but also enhances the synergy between them, resulting in a sophisticated spoken chatbot capable of delivering high-quality, personalized interactions.
        \begin{equation}
        \begin{aligned}
        Loss = L_{LLM-Chat}+L_{LLM-TTS}
        \end{aligned}
        \end{equation}

\section{Experiments}
    In our evaluation of JoyTTS's voice cloning capabilities, we utilized the SM (Similarity Measure) and WER (Word Error Rate) metrics on the seed-tts-zh dataset. These metrics are critical in assessing the system's performance in replicating the nuances of the original speaker's voice and ensuring the intelligibility and accuracy of the generated speech. The results indicate that JoyTTS excels in producing high-quality, natural-sounding speech while maintaining precise voice cloning. This showcases its potential for applications requiring personalized and realistic voice synthesis.

    Furthermore, JoyTTS demonstrates impressive computational efficiency, achieving a latency of just 1.8 seconds on a single NVIDIA 4090D, without any engineering optimizations. This low latency suggests that JoyTTS can deliver real-time interactions, making it suitable for dynamic environments where quick response times are crucial. Future optimizations could further reduce this latency, enhancing its applicability in various real-world scenarios. Overall, these results underscore JoyTTS's effectiveness and efficiency in voice cloning and speech synthesis.

    \begin{table*}[!hb]
    \centering
    \renewcommand\arraystretch{1.3}
    \setlength{\tabcolsep}{1.5mm}
    \caption{Speaker similarity (SS), content consistency (WER) results on SEED test-zh.}    
    \begin{tabular}{cccccccc}
        \toprule
        Model     & SS $\uparrow$  & WER $\downarrow$   \\
        gpt-sovits     & 0.55 & 5.13  \\
        cosyvoice2  & \textbf{0.748}          & \textbf{1.45}  \\ 
        JoyTTS     & 0.73 & 5.09  \\
        \bottomrule
    \end{tabular}
    \label{tab:1}
    \end{table*}

\section{Conclusions}
    JoyTTS  integrate sophisticated LLM (Large Language Model) and TTS (Text-to-Speech) models to enable voice cloning. This integration is achieved through the innovative use of hidden states as an intermediary layer, which not only enhances the model's performance but also reduces the latency typically associated with cloned dialogues. By effectively bridging the gap between text generation and speech synthesis, JoyTTS delivers a seamless and coherent conversational experience.

    The open-source nature of the project invites further research and development, encouraging enhancements in voice interaction systems. This openness fosters collaboration and innovation, allowing researchers and developers to build upon the existing framework to explore new techniques and applications in the realm of spoken AI.

    Looking ahead, one promising area of development is the introduction of emotion control inputs within the LLM. This would enable the automatic modulation of the TTS output to reflect the optimal emotional tone in conversations. By incorporating emotional intelligence into the model, JoyTTS could achieve more natural and engaging interactions, further enhancing user experience and satisfaction.

    In summary, JoyTTS aims to contribute to the ongoing development of chatbot capabilities by exploring new possibilities in emotionally aware and contextually relevant voice interaction systems. We hope that its contributions to the field will help unlock the potential for AI-driven communication tools to become more personalized and effective across various applications.
{\small
\bibliographystyle{unsrt}
\bibliography{references}
}

\end{document}